\documentclass[journal]{IEEEtran}

\usepackage{amsmath,amsfonts,amsthm,amssymb}
\usepackage{bbm}
\usepackage{graphicx}
\usepackage{cite}
\usepackage{color}
\usepackage{epstopdf}
\usepackage{float}
\usepackage{subfigure}
\usepackage{setspace}
\usepackage{stfloats}
\usepackage{url}

\def\BibTeX{{\rm B\kern-.05em{\sc i\kern-.025em b}\kern-.08em
    T\kern-.1667em\lower.7ex\hbox{E}\kern-.125emX}}

\begin{document}

\title{Cooperative Double-IRS Aided Communication: Beamforming Design and Power Scaling}

\author{Yitao~Han,~\IEEEmembership{Student~Member,~IEEE,}
	Shuowen~Zhang,~\IEEEmembership{Member,~IEEE,}\\
        Lingjie~Duan,~\IEEEmembership{Senior~Member,~IEEE,}
        and~Rui~Zhang,~\IEEEmembership{Fellow,~IEEE}
\thanks{Y.~Han and L.~Duan are with the Engineering Systems and Design Pillar, Singapore University of Technology and Design (e-mail: yitao\_han@mymail.sutd.edu.sg, lingjie\_duan@sutd.edu.sg), Y.~Han is also with the Department of Electrical and Computer Engineering, National University of Singapore. L.~Duan is the corresponding author.}
\thanks{S.~Zhang and R.~Zhang are with the Department of Electrical and Computer Engineering, National University of Singapore (e-mail: elezhsh@nus.edu.sg, elezhang@nus.edu.sg).}
}

\maketitle

\begin{abstract}

Intelligent reflecting surface (IRS) is a promising technology to support high performance wireless communication. By adaptively configuring the reflection amplitude and/or phase of each passive reflecting element on it, the IRS can reshape the electromagnetic environment in favour of signal transmission. This letter advances the existing research by proposing and analyzing a \emph{double-IRS} aided wireless communication system. Under the reasonable assumption that the reflection channel from IRS $1$ to IRS $2$ is of rank $1$ (e.g., line-of-sight channel), we propose a joint passive beamforming design for the two IRSs. Based on this, we show that deploying two cooperative IRSs with in total $K$ elements can yield a power gain of order $\mathcal{O}(K^4)$, which greatly outperforms the case of deploying one traditional IRS with a power gain of order $\mathcal{O}(K^2)$. Our simulation results validate that the performance of deploying two cooperative IRSs is significantly better than that of deploying one IRS given a sufficient total number of IRS elements. We also extend our line-of-sight channel model to show how different channel models affect the performance of the double-IRS aided wireless communication system.

\end{abstract}

\begin{IEEEkeywords}

Intelligent reflecting surface, channel modelling, passive beamforming, power scaling.

\end{IEEEkeywords}

\section{Introduction}

With the recent technology advances in micro electromechanical systems (MEMS) and metamaterial \cite{cui2014coding}, controlling the amplitude and/or phase of the reflected signal in real time via a programmable surface becomes feasible. This enables an innovative wireless device and network component---the intelligent reflecting surface (IRS). An IRS usually consists of a large number of passive reflecting elements, and by adaptively configuring the reflection amplitude and/or phase of each element, the IRS can modify the wireless propagation environment to fit specific needs \cite{wu2019towards}. Different from the traditional active relay, the IRS only leverages passive reflection, which does not require expensive hardware or high energy consumption \cite{wu2019towards}. Current research on IRS mainly focuses on passive or joint beamforming design \cite{basar2019wireless,wu2019intel,zhang2019capacity,huang2019recon,yang2019intel,tan2018enabling,hu2018beyond}, which only study the scenario with one IRS or multiple faraway IRSs each independently serving its associated users in the vicinity, thus no cooperation or joint passive beamforming among multiple IRSs is considered, to the authors' best knowledge.

In this letter, we make the first attempt to study a double-IRS aided wireless communication system as illustrated in Fig.~\ref{sm}, where a user is served by the base station (BS) through the \emph{double reflection link} (BS-IRS$1$-IRS$2$-user), while other links are unavailable due to severe blockage. Generally speaking, it is difficult to design the passive beamformers of the two IRSs for aligning the channel from IRS $1$ to IRS $2$. Here by carefully planning two IRSs' locations, we reasonably assume that channel is line-of-sight (LoS), and prove that deploying two cooperative IRSs with in total $K$ elements can yield a power gain of order $\mathcal{O}(K^4)$, which greatly outperforms deploying one traditional IRS with a power gain of order $\mathcal{O}(K^2)$ in the user's vicinity (i.e., all $K$ elements are deployed on IRS $2$ in Fig.~\ref{sm})\cite{wu2019towards}. We model the LoS channel between the two IRSs based on their geometric relationship, and jointly design their passive beamformers to achieve the aforementioned power gain. Our simulation results validate the above theoretical power scaling as well as performance comparison. We also extend our LoS channel model to the more general Rician fading channel model and examine how different channel models affect the system performance.

\section{System Model}

\begin{figure}[t]
\centering
\includegraphics[width=0.39\textwidth]{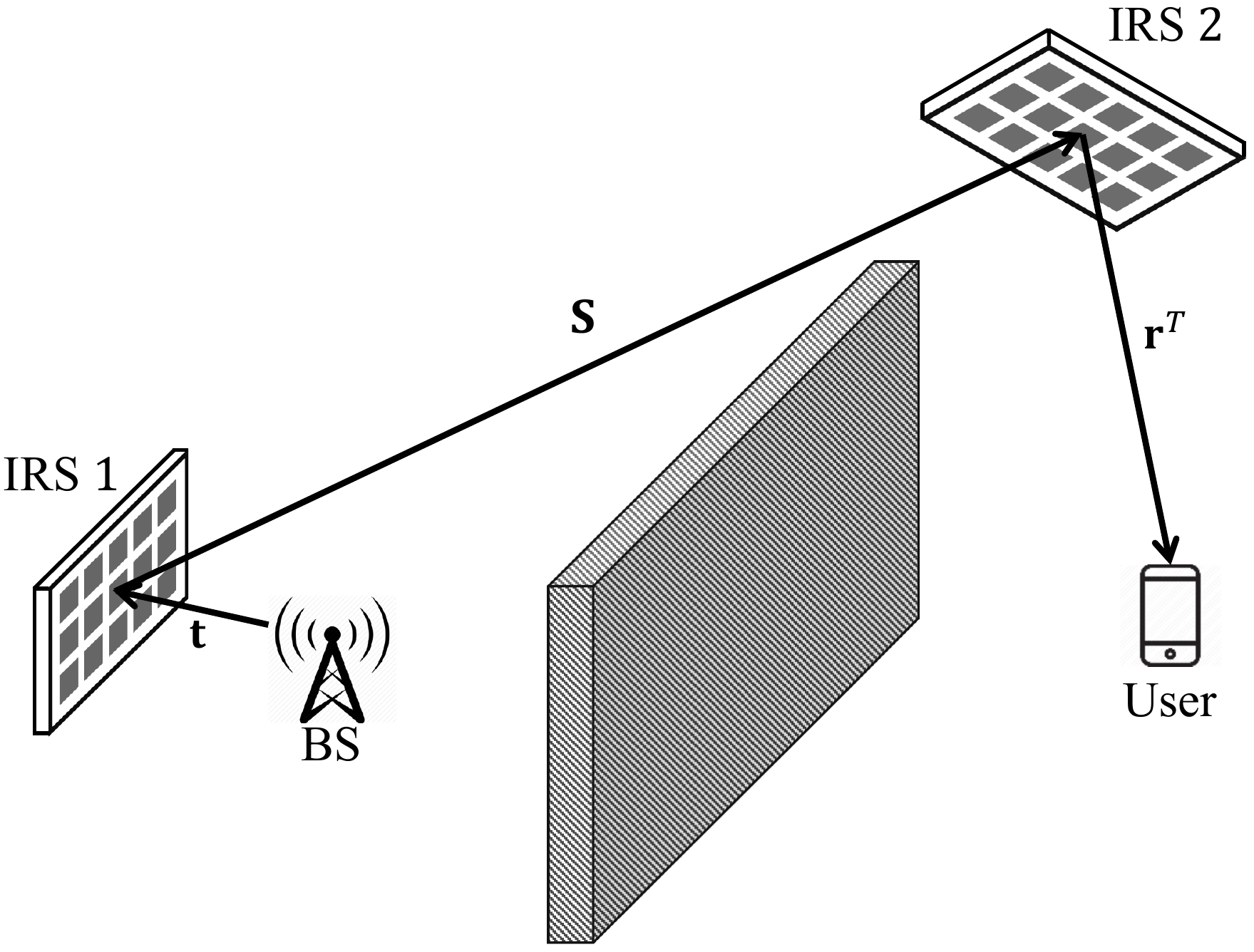}
\caption{A wireless communication system aided by two cooperative IRSs, where the user is served by the BS through the double reflection link (BS-IRS$1$-IRS$2$-user), while other links are unavailable due to severe blockage.}
\label{sm}
\vspace{-1em}
\end{figure}

In this letter, we aim to study the effect of deploying two cooperative IRSs for improving the performance of a wireless communication system. As shown in Fig.~\ref{sm}, we consider the downlink communication from the BS to the user, both of which are assumed to be equipped with a single antenna.\footnote{We consider the downlink case for the purpose of exposition, while all the results are directly applicable to the uplink case.} We denote the location of the BS as $\mathbf{u}_1\in \mathbb{R}^{3\times 1}$ and that of the user as $\mathbf{u}_2\in \mathbb{R}^{3\times 1}$ under a three-dimensional (3D) Cartesian coordinate system. We focus on a challenging scenario where the direct link from the BS to the user is blocked by obstacles. To enhance the communication, we place IRS $1$ near to the BS and IRS $2$ near to the user, such that the user can be served by the BS through the double reflection link, i.e., BS-IRS$1$-IRS$2$-user. There are totally $K$ passive reflecting elements as our budget, and IRS $i$ has $K_i$ elements with $\sum_{i\in\{1,2\}}K_i=K$. We further define $\mathcal{K}_i$ as the set containing all the elements on IRS $i$.

We denote $\mathbf{t}\in\mathbb{C}^{K_1\times1}$, $\mathbf{S}\in\mathbb{C}^{K_2\times K_1}$, and $\mathbf{r}^T\in\mathbb{C}^{1\times K_2}$ as the baseband equivalent channels from BS to IRS $1$, from IRS $1$ to IRS $2$, and from IRS $2$ to user, respectively, where $(\cdot)^T$ is the transpose operation. We carefully deploy the two IRSs such that $\mathbf{t}$, $\mathbf{S}$, and $\mathbf{r}^T$ can be modelled as LoS channels, while other channels are negligible due to severe blockage. Later in Section IV, we will examine how different channel models affect the performance. Each IRS is connected to a controller, which can adjust the reflection coefficients of its elements, i.e., $\phi_{i,k_i}$ for $i\in\{1,2\}$ and $k_i\in\mathcal{K}_i$, to control the direction of the reflected signals. We denote the reflection coefficient matrix of IRS $i$ to be $\mathbf{\Phi}_i=\text{diag}\{\phi_{i,1},\cdots,\phi_{i,k_i},\cdots,\phi_{i,K_i}\}\in\mathbb{C}^{K_{i}\times K_{i}}$. For maximal reflection and ease of practical implementation \cite{wu2019towards}, we further set $|\phi_{i,k_i}|=1$ for $i\in\{1,2\}$ and $k_i\in\mathcal{K}_i$, and assume that IRS $1$ is oriented towards the BS (i.e., the signals from the BS arrive at IRS $1$ perpendicularly). With the above setup, the effective channel from BS to user is modelled as
\begin{equation}
\begin{aligned}
h = \mathbf{r}^T\mathbf{\Phi}_2\mathbf{S}\mathbf{\Phi}_1\mathbf{t}.
\label{eq1}
\end{aligned}
\end{equation}

In the following, we provide the exact characterization of $\mathbf{t}$, $\mathbf{S}$, and $\mathbf{r}^T$. Denote the distance between BS and element $k_{1}$ on IRS $1$ as $d_{k_1,\mathbf{u}_1}$, the distance between element $k_1$ on IRS $1$ and element $k_2$ on IRS $2$ as $d_{k_2,k_1}$, and the distance between element $k_2$ on IRS $2$ and user as $d_{\mathbf{u}_2,k_2}$, respectively. Under the LoS assumption, the entry in row $k_1$ of the channel from BS to IRS $1$, $\mathbf{t}$, is given by
\begin{equation}
\begin{aligned}
(\mathbf{t})_{k_1}=\frac{\sqrt{\alpha}}{d_{k_1,\mathbf{u}_1}}\exp\bigg(\frac{-j2\pi}{\lambda}d_{k_1,\mathbf{u}_1}\bigg),\ \ k_1\in\mathcal{K}_1,
\label{eq2}
\end{aligned}
\end{equation}
where $\alpha$ is the channel power gain at the reference distance $d_{\text{ref}}=1$ meter (m), and $\lambda$ is the carrier wavelength. Similarly, the entry in row $k_2$ and column $k_1$ of the channel from IRS $1$ to IRS $2$, $\mathbf{S}$, is
\begin{equation}
\begin{aligned}
(\mathbf{S})_{k_2,k_1}\!=\!\frac{\sqrt{\alpha}}{d_{k_2,k_1}}\!\exp\!\bigg(\!\frac{-j2\pi}{\lambda}d_{k_2,k_1}\!\bigg),\ k_1\in\mathcal{K}_1, k_2\in\mathcal{K}_2.
\label{eq3}
\end{aligned}
\end{equation}Finally, the entry in column $k_2$ of the channel from IRS $2$ to user, $\mathbf{r}^T$, is
\begin{equation}
\begin{aligned}
(\mathbf{r}^T)_{k_2}=\frac{\sqrt{\alpha}}{d_{\mathbf{u}_2,k_2}}\exp\bigg(\frac{-j2\pi}{\lambda}d_{\mathbf{u}_2,k_2}\bigg),\ \ k_2\in\mathcal{K}_2.
\label{eq4}
\end{aligned}
\end{equation}

In order to design the reflection coefficient matrices $\mathbf{\Phi}_1$ and $\mathbf{\Phi}_2$ in \eqref{eq1} for achieving the optimal performance, we assume that the elements on both IRSs are equipped with receive RF chain \cite{wu2019towards}. Therefore, the two IRSs can directly obtain the channel state information (CSI) of the channel from BS to IRS $1$, $\mathbf{t}$, and the channel from IRS $2$ to user, $\mathbf{r}^T$, via channel estimation. However, it is difficult to directly estimate the channel from IRS $1$ to IRS $2$, $\mathbf{S}$, because both sides of the channel are passive reflecting IRSs without the capability of transmitting pilot signals for channel estimation. To tackle this issue, we propose an efficient approach for deriving the channel between the two IRSs in the next section. Based on this, we then jointly design the reflection coefficient matrices $\mathbf{\Phi}_1$ and $\mathbf{\Phi}_2$.

\section{Passive Beamforming Design for Cooperative IRSs}

\begin{figure}[t]
\centering
\includegraphics[width=0.3\textwidth]{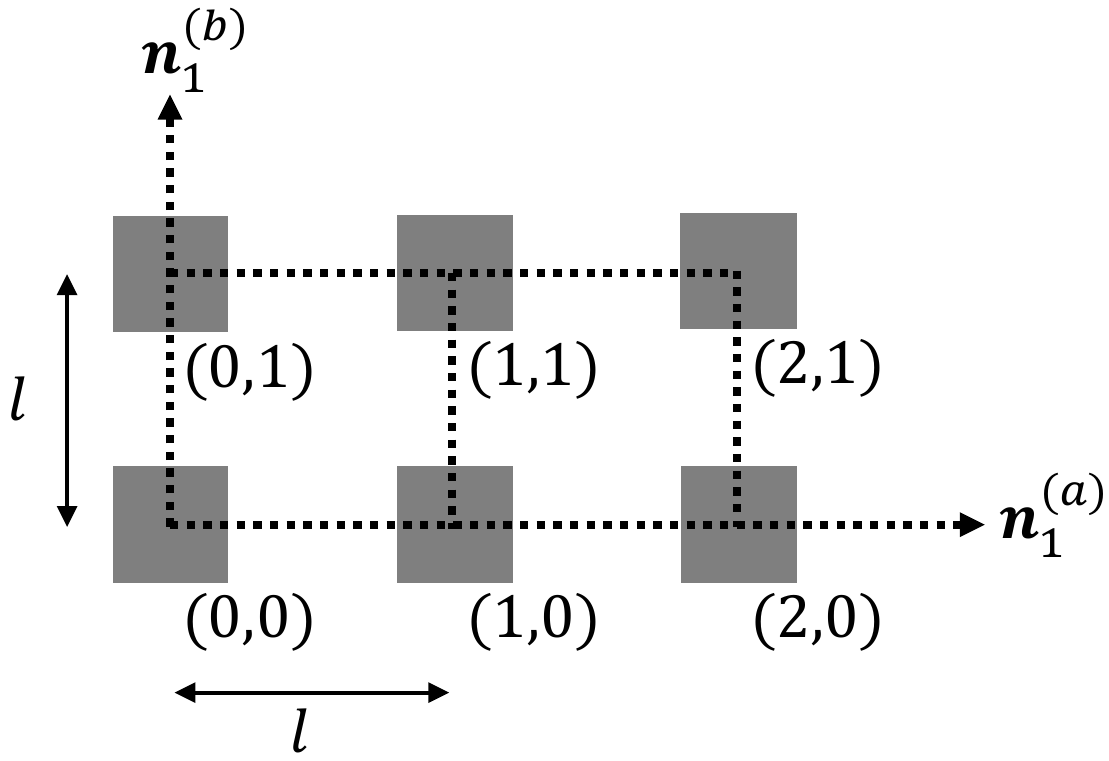}
\caption{An example of a rectangular IRS $1$ with $K_1=6$ elements. Here $K_1^{(a)}=3$ elements in direction $\mathbf{n}_1^{(a)}$ and $K_1^{(b)}=2$ elements in direction $\mathbf{n}_1^{(b)}$.}
\label{irs}
\vspace{-1em}
\end{figure}

In this section, we first provide a tractable approach to characterize the LoS channel from IRS $1$ to IRS $2$, $\mathbf{S}$, based on their geometric relationship, then we propose the joint passive beamforming design for the two IRSs.

\subsection{Tractable Characterization of Inter-IRS Channel}

Without loss of generality, we assume that the passive reflecting elements, both on IRS $1$ and IRS $2$, are arranged in a rectangular shape, see an example of IRS $1$ as shown in Fig.~\ref{irs}. The elements are placed on lines along two orthogonal base directions $\mathbf{n}_{1}^{(a)}\in \mathbb{R}^{3\times 1}$ and $\mathbf{n}_{1}^{(b)}\in \mathbb{R}^{3\times 1}$, with $\|\mathbf{n}_{1}^{(a)}\|=\|\mathbf{n}_{1}^{(b)}\|=1$, where $\|\cdot\|$ is the Euclidean norm of a vector. Any two adjacent elements in each base direction have a uniform separation distance $l$. The numbers of elements in IRS $1$'s first and second base directions are denoted by $K_{1}^{(a)}$ and $K_{1}^{(b)}$, respectively, where $K_1=K_{1}^{(a)}K_{1}^{(b)}$. We denote the position of any particular element $k_1$ on IRS $1$ as $(k_{1}^{(a)},k_{1}^{(b)})$, which summarizes the indices in IRS $1$'s first and second base directions, where $k_{1}^{(a)}\in\{0,\cdots,K_{1}^{(a)}-1\}$ and $k_{1}^{(b)}\in\{0,\cdots,K_{1}^{(b)}-1\}$. Thus, we have the unique mapping between $k_1$ and $(k_{1}^{(a)},k_{1}^{(b)})$, i.e., $k_1=k_{1}^{(a)}+1+k_{1}^{(b)}K_{1}^{(a)}.$
We further denote the location of element $(0,0)$ on IRS $1$ as $\mathbf{v}_1\in \mathbb{R}^{3\times 1}$. Similarly, we denote $\mathbf{n}_{2}^{(a)}$, $\mathbf{n}_{2}^{(b)}$, $K_{2}^{(a)}$, $K_{2}^{(b)}$, $k_{2}^{(a)}$, $k_{2}^{(b)}$, $\mathbf{v}_2$ for IRS $2$, and we have $k_2=k_{2}^{(a)}+1+k_{2}^{(b)}K_{2}^{(a)}$. Note that $k_1$ and $(k_{1}^{(a)},k_{1}^{(b)})$ are interchangeable for the rest of the letter, so are $k_2$ and $(k_{2}^{(a)},k_{2}^{(b)})$.

\begin{figure}[t]
\centering
\includegraphics[width=0.45\textwidth]{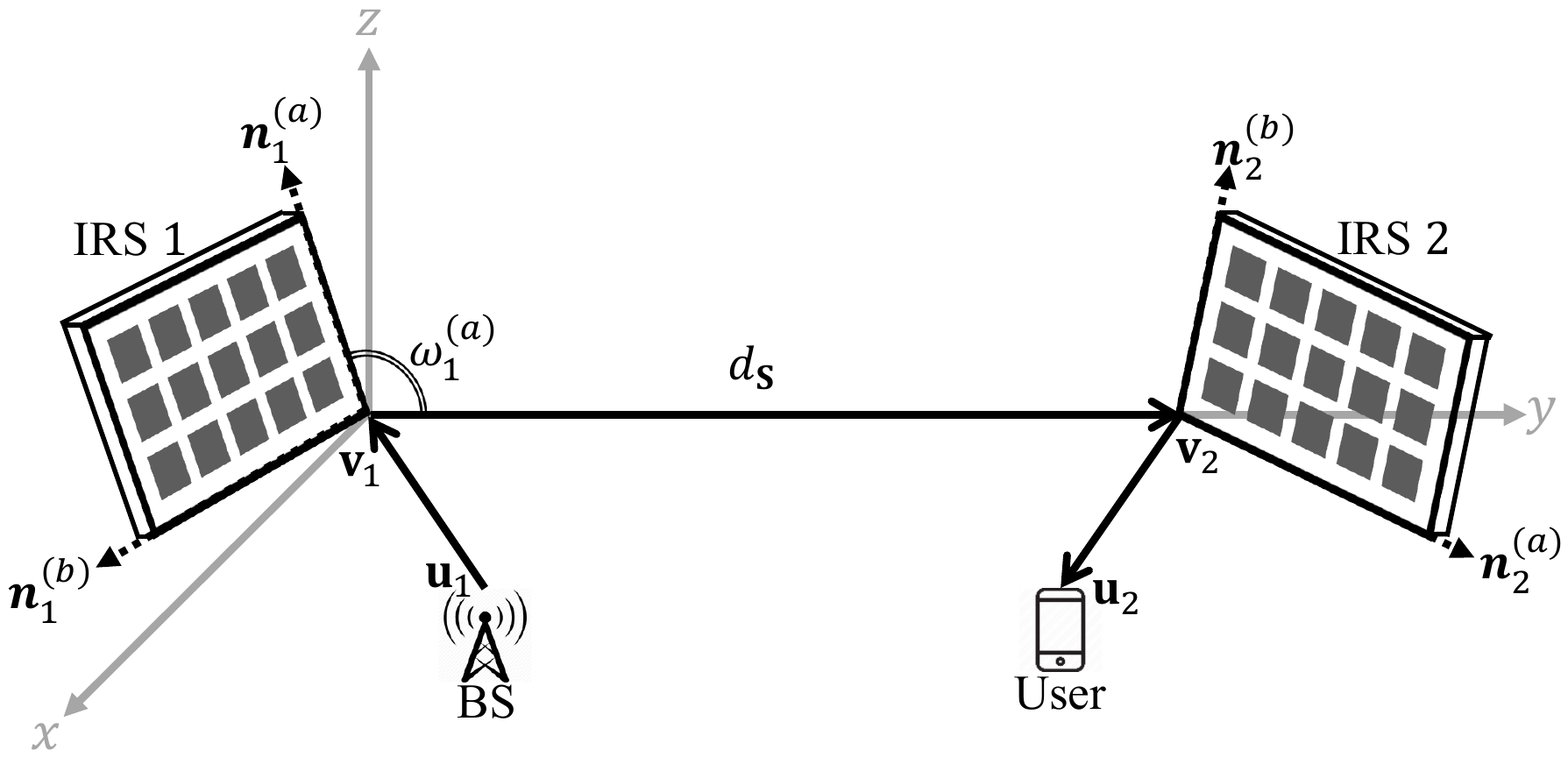}
\caption{Two IRSs in 3D space (the sizes of IRSs are exaggerated).}
\label{3dmodel}
\vspace{-0.7em}
\end{figure}

As illustrated in Fig.~\ref{3dmodel}, we normalize the location of IRS $1$'s element $(0,0)$ to be $\mathbf{v}_1=[0,0,0]^T$. Similarly, the location of IRS $2$'s element $(0,0)$ is purposely set as $\mathbf{v}_2=[0,d_{\mathbf{S}},0]^T$, where $d_{\mathbf{S}}=\|\mathbf{v}_2-\mathbf{v}_1\|$ denotes the distance between element $(0,0)$ on IRS $1$ and element $(0,0)$ on IRS $2$.

According to \cite{gesbert2002outdoor} and \cite{larsson2005lattice}, the LoS channel from IRS $1$ to IRS $2$, $\mathbf{S}$, can be assumed to be a \emph{rank-one} matrix if 
\begin{equation}
\begin{aligned}
d_{\mathbf{S}}\gg\frac{\sqrt{K_2}l^2}{\lambda}.
\label{add}
\end{aligned}
\end{equation}
In this letter, we assume that \eqref{add} holds due to the generally large distance between the two IRSs as compared to their sizes. In the following, we present the characterization of $\mathbf{S}$ as a rank-one matrix based on the geometric relationship between the two IRSs.

We denote $\omega_1^{(a)}=\arccos\Big(\frac{(\mathbf{v}_2-\mathbf{v}_1)\cdotp\mathbf{n}_1^{(a)}}{\|\mathbf{v}_2-\mathbf{v}_1\|\cdotp\|\mathbf{n}_1^{(a)}\|}\Big)\in[0,\pi]$ as the angle between $(\mathbf{v}_2-\mathbf{v}_1)$ and $\mathbf{n}_1^{(a)}$, as shown in Fig.~\ref{3dmodel}, where $\cdotp$ is the dot product. Similarly, we denote $\omega_1^{(b)}$, $\omega_2^{(a)}$, and $\omega_2^{(b)}$ as the angles between $(\mathbf{v}_2-\mathbf{v}_1)$ and $\mathbf{n}_1^{(b)}$, between $(\mathbf{v}_2-\mathbf{v}_1)$ and $\mathbf{n}_2^{(a)}$, and between $(\mathbf{v}_2-\mathbf{v}_1)$ and $\mathbf{n}_2^{(b)}$, respectively. Note that the location of element $(k_i^{(a)},k_i^{(b)})$ on IRS $i$ is $\mathbf{v}_i+\!k_{i}^{(a)}l\mathbf{n}_{i}^{(a)}\!+\!k_{i}^{(b)}l\mathbf{n}_{i}^{(b)}$, therefore the distance between element $(k_1^{(a)},k_1^{(b)})$ on IRS $1$ and element $(k_2^{(a)},k_2^{(b)})$ on IRS $2$ is
\begin{equation}
\begin{aligned}
&d_{k_{2}\!,k_{1}}\!=\!\Big\|\mathbf{v}_2\!+\!k_{2}^{(a)}l\mathbf{n}_{2}^{(a)}\!\!+\!k_{2}^{(b)}l\mathbf{n}_{2}^{(b)}\!\!-\!\mathbf{v}_1\!-\!k_{1}^{(a)}l\mathbf{n}_{1}^{(a)}\!\!-\!k_{1}^{(b)}l\mathbf{n}_{1}^{(b)}\Big\|\\
&=\!\Big(\big(d_{\mathbf{S}}\!+\!k_{2}^{(a)}l\cos(\omega_2^{(a)})\!+\!k_{2}^{(b)}l\cos(\omega_2^{(b)})\!-\!k_{1}^{(a)}l\cos(\omega_1^{(a)})\nonumber
\end{aligned}
\end{equation}
\vspace{-0.5em}
\begin{equation}
\begin{aligned}
\!\!\!\!\!\!\!\!\!\!\!\!\!\!\!\!\!\!\!\!\!\!\!\!\!\!\!\!-k_{1}^{(b)}l\cos(\omega_1^{(b)})\big)^2+\delta_{\mathbf{x}}^2+\delta_{\mathbf{z}}^2\Big)^{1/2}
\label{add1}
\end{aligned}
\end{equation}
\begin{equation}
\begin{aligned}
\!\!\!\!\!\!\!\!\!\!\!\approx\!d_{\mathbf{S}}\!+\!k_{2}^{(a)}l\cos(\omega_2^{(a)})\!+\!k_{2}^{(b)}l\cos(\omega_2^{(b)})\!-\!k_{1}^{(a)}l\cos(\omega_1^{(a)})\nonumber
\end{aligned}
\end{equation}
\begin{equation}
\begin{aligned}
\!\!\!\!\!\!\!\!\!\!\!\!\!\!\!\!\!\!\!\!\!\!\!\!\!\!\!\!\!\!\!\!\!\!\!\!\!\!\!\!\!\!\!\!\!\!\!\!\!\!\!\!\!\!\!\!\!\!\!\!\!-k_{1}^{(b)}l\cos(\omega_1^{(b)}),
\label{eq7}
\end{aligned}
\end{equation}where $\delta_{\mathbf{x}}$ and $\delta_{\mathbf{z}}$ are the distances in $\mathbf{x}$ and $\mathbf{z}$ directions related to the sizes of IRSs. Recall that we assume $d_{\mathbf{S}}$ is much larger than the sizes of IRSs, we can therefore accurately approximate \eqref{add1} by omitting $\delta_{\mathbf{x}}$ and $\delta_{\mathbf{z}}$ as \eqref{eq7}.

Based on the simplification in \eqref{eq7}, we can approximate the channel entry from element $(k_1^{(a)},k_1^{(b)})$ on IRS $1$ to element $(k_2^{(a)},k_2^{(b)})$ on IRS $2$ in \eqref{eq3} as
\begin{equation}
\begin{aligned}
(\mathbf{S})_{k_2,k_1}\!\!\approx\!\!\frac{\sqrt{\alpha}}{d_{\mathbf{S}}}\!\exp\!\bigg(\!\frac{-j2\pi }{\lambda}\!\Big(\!d_{\mathbf{S}}\!+\!k_{2}^{(a)}l\cos(\omega_2^{(a)}\!)\!+\!k_{2}^{(b)}l\cos(\omega_2^{(b)}\!)\nonumber
\end{aligned}
\end{equation}
\vspace{-0.5em}
\begin{equation}
\begin{aligned}
\ \ \ \ \ \ \ \ \ \ \ \ \ \ \ \ \ \ \ \ \ -k_{1}^{(a)}l\cos(\omega_1^{(a)})\!-\!k_{1}^{(b)}l\cos(\omega_1^{(b)})\Big)\bigg).
\label{eq8}
\end{aligned}
\end{equation}
Similarly due to the large $d_{\mathbf{S}}$ as compared to the sizes of IRSs, the difference between $\{d_{k_2,k_1}\}$ has negligible effect on the path loss $\sqrt{\alpha}/d_{k_2,k_1}$ in \eqref{eq3}, thus we further use $\sqrt{\alpha}/d_{\mathbf{S}}$ in \eqref{eq8} to approximate the path loss between any element on IRS $1$ and any element on IRS $2$.

We observe from \eqref{eq8} that each channel entry, $(\mathbf{S})_{k_2,k_1}$ for $k_1\in\mathcal{K}_1$ and $k_2\in\mathcal{K}_2$, can be represented as the product of two scalars $({\mathbf{g}}_1)_{k_1}\overset{\Delta}{=}\exp\big(\frac{j2\pi}{\lambda}\big(k_{1}^{(a)}l\cos(\omega_1^{(a)}) + k_{1}^{(b)}l\cos(\omega_1^{(b)})\big)\big)$ and $({\mathbf{g}}_2)_{k_2}\overset{\Delta}{=}\exp\big(\frac{-j2\pi}{\lambda}\big(k_{2}^{(a)}l\cos(\omega_2^{(a)}) + k_{2}^{(b)}l\cos(\omega_2^{(b)})\big)\big)$. Therefore, we can decompose the channel matrix $\mathbf{S}$ as the product of two signature vectors $\mathbf{g}_1$ and $\mathbf{g}_2$, namely,
\begin{equation}
\begin{aligned}
\mathbf{S} \approx \frac{\sqrt{\alpha}}{d_{\mathbf{S}}}\exp\bigg(\frac{-j2\pi d_{\mathbf{S}}}{\lambda}\bigg){\mathbf{g}}_2{\mathbf{g}}_1^T,
\label{eq11}
\end{aligned}
\end{equation}where ${\mathbf{g}}_1=\big[({\mathbf{g}}_1)_{1},\!\cdots\!,({\mathbf{g}}_1)_{k_1},\!\cdots\!,({\mathbf{g}}_1)_{K_1}\big]^T$ and ${\mathbf{g}}_2=\big[({\mathbf{g}}_2)_{1},\!\cdots\!,({\mathbf{g}}_2)_{k_2},\!\cdots\!,({\mathbf{g}}_2)_{K_2}\big]^T$.

With \eqref{eq11}, we can calculate the CSI of the channel from IRS $1$ to IRS $2$, $\mathbf{S}$, based on the geometric relationship between the two IRSs, i.e., $\{\mathbf{v}_{1},\mathbf{v}_{2},\mathbf{n}_{1}^{(a)},\mathbf{n}_{1}^{(b)},\mathbf{n}_{2}^{(a)},\mathbf{n}_{2}^{(b)}\}$.

\subsection{Design of Joint Passive Beamforming}

In general, it is difficult to design the reflection coefficient matrices $\mathbf{\Phi}_1$ and $\mathbf{\Phi}_2$ in \eqref{eq1} for aligning all the entries in $\mathbf{S}$, i.e., $\{(\mathbf{S})_{k_2,k_1}\}$, especially when $\mathbf{S}$ is of high rank (e.g.,
Rayleigh fading channel). However, under our considered LoS channel model, $\mathbf{S}$ in \eqref{eq11} is of rank $1$, which means that all the entries $\{(\mathbf{S})_{k_2,k_1}\}$ are highly correlated. Therefore, we can obtain the following approximated passive beamforming design that aligns all the entries in $\mathbf{S}$, thus achieves a power gain scalable with $K_1$ and $K_2$.

By assuming the channel from IRS $1$ to IRS $2$ is LoS and the two IRSs are sufficiently faraway, we configure the reflection coefficient matrix of IRS $1$ to
\begin{equation}
\begin{aligned}
\phi_{1,k_1}=\bigg(\frac{({\mathbf{g}}_1)_{k_1}(\mathbf{t})_{k_1}}{|(\mathbf{t})_{k_1}|}\bigg)^*, \ \ k_1\in\mathcal{K}_1,
\label{eq12}
\end{aligned}
\end{equation}
and that of IRS $2$ to
\begin{equation}
\begin{aligned}
\phi_{2,k_2}=\bigg(\frac{(\mathbf{r}^T)_{k_2}({\mathbf{g}}_2)_{k_2}}{|(\mathbf{r}^T)_{k_2}|}\bigg)^*, \ \ k_2\in\mathcal{K}_2,
\label{eq13}
\end{aligned}
\end{equation}
where $(\cdot)^*$ is the conjugate operation.

IRS $1$ with the passive beamformer in \eqref{eq12} reflects the signals from the BS and then beams them towards element $(0,0)$ on IRS $2$. Given the channel decomposition in \eqref{eq11}, we can easily see that it automatically yields a $K_1^2$-fold power gain at each element on IRS $2$. IRS $2$ with the passive beamformer in \eqref{eq13} further reflects the signals from IRS $1$ and beams them towards the user, therefore
the user can enjoy a $(K_1K_2)^2$-fold power gain, i.e.,
\begin{equation}
\begin{aligned}
|h|^2 = \big|\mathbf{r}^T{\mathbf{\Phi}}_2\mathbf{S}{\mathbf{\Phi}}_1\mathbf{t}\big|^2 \approx \frac{\alpha^{3}}{(d_{\mathbf{r}}d_{\mathbf{S}}d_{\mathbf{t}})^2}(K_1K_2)^2,
\label{eq14}
\end{aligned}
\end{equation}
where $d_{\mathbf{t}}=\|\mathbf{v}_1-\mathbf{u}_1\|$ is the distance between BS and element $(0,0)$ on IRS $1$, and $d_{\mathbf{r}}=\|\mathbf{u}_2-\mathbf{v}_2\|$ is the distance between element $(0,0)$ on IRS $2$ and user, i.e., we use $\sqrt{\alpha}/d_{\mathbf{t}}$ to approximate the path loss between BS and any element on IRS $1$ in \eqref{eq2}, and $\sqrt{\alpha}/d_{\mathbf{r}}$ to approximate the path loss between any element on IRS $2$ and user in \eqref{eq4}.

In practice, we cannot change the numbers of elements $K_1$ and $K_2$ on IRSs once the IRSs are deployed. Thus, we are interested in finding the optimal number of elements on each IRS to maximize \eqref{eq14}, and we have the following proposition.

\indent \underline{\emph{Proposition 3.1:}} \ By optimally equipping IRS $1$ and IRS $2$ with the same number of elements, i.e., $K_1=K_2=K/2$, deploying two cooperative IRSs with the refection coefficient matrices given in \eqref{eq12} and \eqref{eq13} can lead to a $K^4$-fold power gain, i.e.,
\begin{equation}
\begin{aligned}
|h|_{\text{opt}}^2\approx\frac{\alpha^{3}}{(4d_{\mathbf{r}}d_{\mathbf{S}}d_{\mathbf{t}})^2}K^4.
\label{eq15}
\end{aligned}
\end{equation}

Note that Proposition 3.1 can be easily proved by taking the first-order derivative of \eqref{eq14} over $K_1$, with $K_2=K-K_1$. This is much more promising than the $K^2$-fold power gain brought by a single IRS near to the user (say, equipping all $K$ elements to IRS $2$ in Fig.~\ref{sm}) \cite{wu2019intel}. It is also worth noting that the enhanced power gain of our proposed scheme relies on the rank-one condition of the inter-IRS channel, i.e., the channel needs to be LoS, and this holds when the inter-IRS distance is sufficiently large. However, such condition may be difficult to satisfy for the general case with more than two IRSs, thus calling for more complicated beamforming and IRS deployment design is left as our future work.

\section{Simulation Results}

In this section, we present simulation results by considering the following setup. We set the locations of the BS, the user, (element $(0,0)$ on) IRS $1$, and (element $(0,0)$ on) IRS $2$ to be $\mathbf{u}_1=[0.87,0.50,0]^T$, $\mathbf{u}_2=[13,92.50,0]^T$, $\mathbf{v}_1=[0,0,0]^T$, and $\mathbf{v}_2=[0,100,0]^T$, respectively. Therefore, the distance between BS and IRS $1$ is $d_{\mathbf{t}}=1$ m, the distance between IRS $1$ and IRS $2$ is $d_{\mathbf{S}}=100$ m, and the distance between IRS $2$ and user is $d_{\mathbf{r}}=15$ m. The two base directions of IRS $1$ are $\mathbf{n}_1^{(a)}=[0,0,1]^T$ and $\mathbf{n}_1^{(b)}=[\sqrt{3}/2,-1/2,0]^T$, hence IRS $1$ faces towards the BS for maximal reflection. While the two base directions of IRS $2$ are $\mathbf{n}_2^{(a)}=[\sqrt{3}/2,1/2,0]^T$ and $\mathbf{n}_2^{(b)}=[0,0,1]^T$. The carrier frequency is set as $5$ GHz, therefore the carrier wavelength is $\lambda=0.06$ m and the channel power gain at the reference distance $d_{\text{ref}}=1$ m is $\alpha=(\lambda/4\pi)^2=-46$ dB. The uniform separation distance between two adjacent IRS elements is set as $l=\lambda/2=0.03$ m. We use the received signal-to-noise ratio (SNR) at the user, i.e., $\Gamma=P|h|^2/\sigma^2$, as the performance metric, with the transmit power of the BS set as $P=43$ dBm and the noise power at the user receiver set as $\sigma^{2}=-60$ dBm.

For performance comparison, we consider a benchmark case with only one IRS, i.e., IRS $2$ is equipped with all $K$ elements. The channel from BS to the single IRS, $\tilde{\mathbf{t}}\in\mathbb{C}^{K\times 1}$, and the channel from the single IRS to user, $\tilde{\mathbf{r}}^T\in\mathbb{C}^{1\times K}$, are assumed to be LoS, following similar definitions as in \eqref{eq2} and \eqref{eq4}, while the direct link from the BS to the user is still unavailable. The single IRS estimates $\tilde{\mathbf{t}}$ and $\tilde{\mathbf{r}}^T$, and employs the reflection coefficient matrix $\tilde{\mathbf{\Phi}}=\text{diag}\{\tilde{\phi}_{1},\cdots,\tilde{\phi}_{k},\cdots,\tilde{\phi}_{K}\}\in\mathbb{C}^{K\times K}$ that aligns all the entries in $\tilde{\mathbf{t}}$ and $\tilde{\mathbf{r}}^T$, i.e.,
\begin{equation}
\begin{aligned}
\tilde{\phi}_{k}=\bigg(\frac{(\tilde{\mathbf{r}}^T)_{k}(\tilde{\mathbf{t}})_{k}}{|(\tilde{\mathbf{r}}^T)_{k}|\cdotp|(\tilde{\mathbf{t}})_{k}|}\bigg)^*, \ \ k\in\mathcal{K}.
\label{eq16}
\end{aligned}
\end{equation}

Since the distance between IRS $1$ and IRS $2$, $d_{\mathbf{S}}$, is very large as compared to that between BS and IRS $1$, $d_{\mathbf{t}}$, in the two-IRS case, we use $d_{\mathbf{S}}$ to approximate the distance between BS and the single IRS, $d_{\tilde{\mathbf{t}}}$, in the one-IRS case, i.e., $d_{\tilde{\mathbf{t}}}\approx d_{\mathbf{S}}$. Note that this gives nearly the same ``product distance'' for the two cases, i.e., $d_{\tilde{\mathbf{r}}}d_{\tilde{\mathbf{t}}}\approx d_{\mathbf{r}}d_{\mathbf{S}}d_{\mathbf{t}}$, where $d_{\mathbf{t}}=1$ m, and $d_{\tilde{\mathbf{r}}}=d_{\mathbf{r}}$ is the distance between the single IRS and user. By doing so, we can focus on the effect of the total number of IRS elements $K$ and fairly compare the two cases. The power gain of the effective channel from BS to user in the one-IRS case is
\begin{equation}
\begin{aligned}
|\tilde{h}|^2 = \big|\tilde{\mathbf{r}}^T\tilde{\mathbf{\Phi}}\tilde{\mathbf{t}}\big|^2\approx\frac{\alpha^2}{(d_{\tilde{\mathbf{r}}}d_{\tilde{\mathbf{t}}})^2}K^2\approx\frac{\alpha^2}{(d_{\mathbf{r}}d_{\mathbf{S}})^2}K^2,
\label{eq17}
\end{aligned}
\end{equation}
which increases with $K^2$. Comparing \eqref{eq15} with \eqref{eq17}, we need at least
\begin{equation}
\begin{aligned}
K=\frac{4}{\sqrt{\alpha}},
\label{rl}
\end{aligned}
\end{equation}
which is around $840$ elements under our simulation setup, to compensate for the extra loss introduced by double reflection, so that the performance of deploying two cooperative IRSs is better than that of deploying one IRS.

\begin{figure}[t]
\centering
\includegraphics[width=0.44\textwidth]{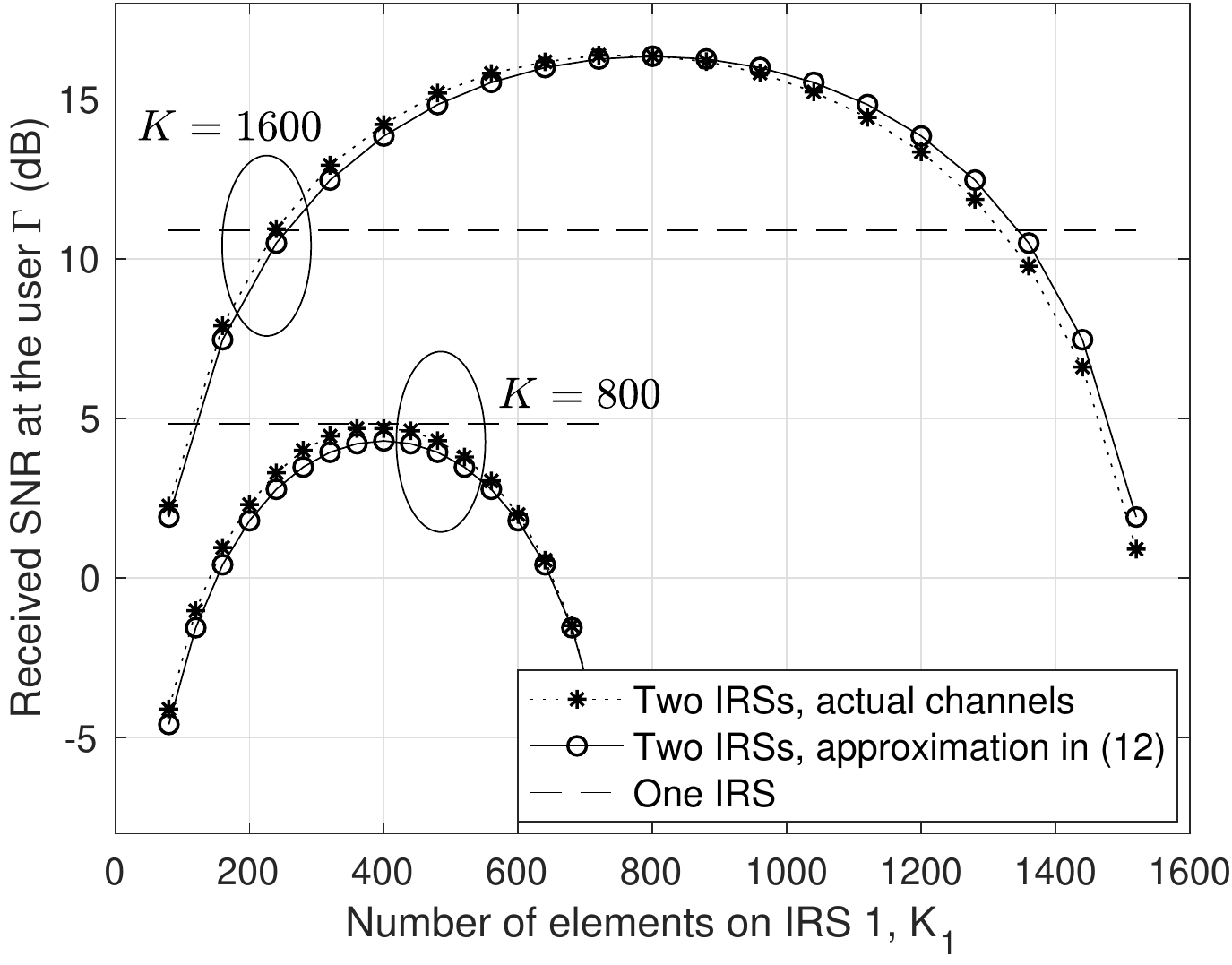}
\vspace{-0.5em}
\caption{Received SNR versus the number of elements on IRS $1$, $K_1$.}
\label{2v1}
\vspace{-1em}
\end{figure}

In Fig.~\ref{2v1}, we plot the received SNR versus the number of elements on IRS $1$, $K_1$, under various total numbers of elements $K$'s for both the two-IRS case and one-IRS case. As we can see, the simulation results based on actual channels match well with the approximation in \eqref{eq14}, indicating the effectiveness of our joint passive beamforming design given in \eqref{eq12} and \eqref{eq13}. The received SNR is maximized when the two IRSs have the same number of elements, i.e., $K_1=K_2=400$ under $K=800$ and $K_1=K_2=800$ under $K=1600$, which is consistent with Proposition 3.1. Note that even with $K_1=K_2=800$ elements, the two IRSs are still at a reasonable size of $800\times l^2=0.72\ \text{m}^2$.

Comparing with the $K^2$-fold power gain brought by deploying one IRS in \eqref{eq17}, although a $K^4$-fold power gain can be achieved by deploying two cooperative IRSs according to \eqref{eq15}, the latter still needs to compensate for the additional loss caused by double reflection, i.e., extra $\alpha$ and $d_{\mathbf{t}}$ in \eqref{eq15}. Therefore, when the total number of elements is not adequate, e.g., $K=800$ in Fig.~\ref{2v1}, to compensate for the aforementioned loss, the performance of deploying two cooperative IRSs is inferior to that of deploying one IRS. But once the total number of elements is large, e.g., $K=1600$ in Fig.~\ref{2v1}, deploying two cooperative IRSs can bring a significant performance gain as compared to deploying one IRS.

Another observation we can make is that by doubling the total number of elements, e.g., from $K=800$ to $K=1600$, the received SNR of the benchmark case with one IRS increases by $\Delta_{\text{I}}\approx 6$ dB, while that of the two-IRS case with $K_1=K_2=K/2$ increases by $\Delta_{\text{II}}\approx 12$ dB. This transfers to $10^{\Delta_{\text{I}}/10}=4$ times the received power increase in the one-IRS case, and $10^{\Delta_{\text{II}}/10}=16$ times the received power increase in the two-IRS case, which verifies our conclusion that the power gain of the one-IRS case is in the order of $\mathcal{O}(K^2)$, while that of the two-IRS case is in the order of $\mathcal{O}(K^4)$.

\begin{figure}[t]
\centering
\includegraphics[width=0.44\textwidth]{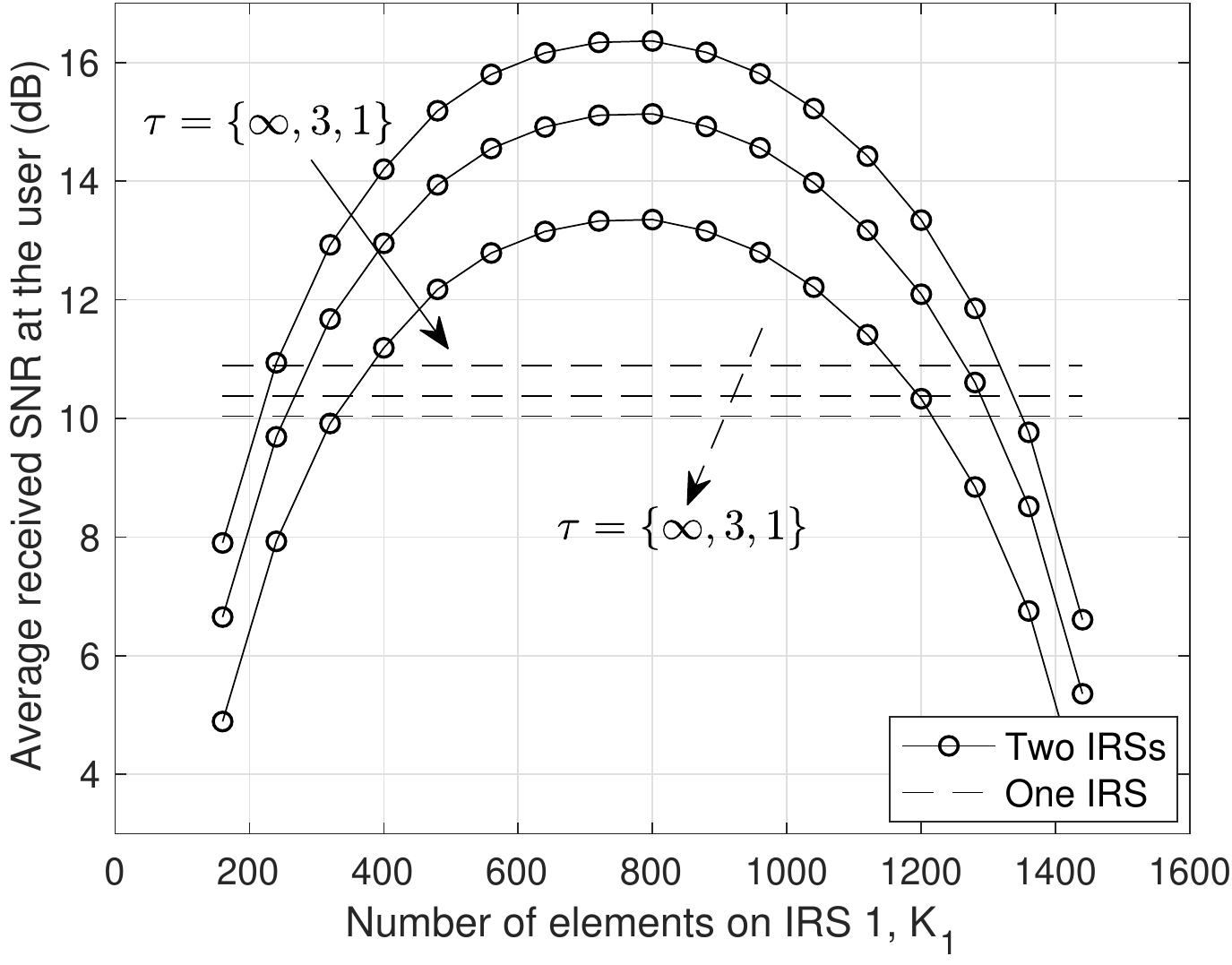}
\vspace{-0.5em}
\caption{Effect of Rician fading factor $\tau$ on the performance.}
\label{rician}
\vspace{-0.6em}
\end{figure}

Finally, we extend our LoS channel model and study the effect of different channel models on the performance of our two-IRS system. We assume that the channel from IRS $1$ to IRS $2$ now follows Rician fading with factor $\tau$, i.e.,
\begin{equation}
\begin{aligned}
\mathbf{S}_{\text{Rician}} = \sqrt{\frac{\tau}{\tau+1}}\mathbf{S}+\sqrt{\frac{1}{\tau+1}}\mathbf{S}_{\text{Rayleigh}},
\label{eq18}
\end{aligned}
\end{equation}where $\mathbf{S}$ is the LoS component as in \eqref{eq3}, and $\mathbf{S}_{\text{Rayleigh}}$ is the scattering component with independent entry $(\mathbf{S}_{\text{Rayleigh}})_{k_2,k_1}\sim\frac{\sqrt{\alpha}}{d_{k_2,k_1}}\mathcal{CN}(0,1)$ for $k_1\in\mathcal{K}_1$ and $k_2\in\mathcal{K}_2$. The channel from BS to IRS $1$, $\mathbf{t}$, and the channel from IRS $2$ to user, $\mathbf{r}^T$, are still assumed to be LoS as in \eqref{eq2} and \eqref{eq4}. IRS $1$ and IRS $2$ employ the passive beamformers in \eqref{eq12} and \eqref{eq13}, respectively. For a fair comparison, in the one-IRS case, the channel from BS to the single IRS, $\tilde{\mathbf{t}}$, also follows Rician fading with the same factor $\tau$ as in \eqref{eq18}, while the channel from the single IRS to user, $\tilde{\mathbf{r}}^T$, is still LoS. The single IRS employs the passive beamformer in \eqref{eq16}. Here, $1,000$ channel realizations are generated to calculate the average received SNR at the user.

In Fig.~\ref{rician}, we plot the average received SNR versus the number of elements on IRS $1$, $K_1$, under different Rician fading factors $\tau=\{\infty,3,1\}$. Note that $\tau=\infty$ stands for LoS channels. The total number of elements is set as $K=1600$. For the two-IRS case, as the Rician fading factor $\tau$ decreases, the average received SNR decreases. This is because our joint passive beamforming design in \eqref{eq12} and \eqref{eq13} is based on the assumption that the channel from IRS $1$ to IRS $2$ is of rank $1$, and if IRS $1$ beams towards one element on IRS $2$, the rest elements on IRS $2$ can enjoy the same power gain. However, the decrease of $\tau$ increases the rank of $\mathbf{S}_{\text{Rician}}$, thus \eqref{eq12} and \eqref{eq13} can no longer align all the entries in $\mathbf{S}_{\text{Rician}}$, which results in performance degradation. While for the one-IRS case, varying the Rician fading factor $\tau$ has little effect on the average received SNR. This is because the passive beamforming design in \eqref{eq16} does not depend on specific channel model but simply aligns all the entries in $\tilde{\mathbf{t}}$ and $\tilde{\mathbf{r}}^T$, whose average magnitudes do not change much with $\tau$. The above performance comparison demonstrates the importance of IRS deployment to create favourable channels for maximizing the joint passive beamforming gain in the multi-IRS system.

\section{Conclusion}

In this letter, we propose and analyze a double-IRS aided wireless communication system. By assuming an LoS channel between the two IRSs, we mathematically characterize the channel based on the IRSs' geometric relationship, and jointly design the passive beamformers for the two cooperative IRSs to achieve a power gain of order $\mathcal{O}(K^4)$, with $K$ denoting the total number of IRS elements. Simulation results validate the performance gain of deploying two cooperative IRSs as compared to placing all $K$ elements on one IRS in the user's vicinity, when $K$ is sufficiently large. We also extend the LoS channel model to the Rician fading channel model, and study the effect of non-LoS propagation on the performance of our proposed two-IRS system and joint passive beamforming design. In future work, it will be interesting to extend this letter to more general setups, such as multi-antenna transmitter/receiver, multiple users, and more (than two) IRSs.

\end{document}